\documentclass[10pt,conference]{IEEEtran}

\usepackage{graphicx}
\usepackage{subfigure}
\usepackage{color}
\usepackage{epsfig}
\usepackage{amssymb}
\usepackage{amsmath}
\usepackage{amsthm}
\usepackage{latexsym}
\usepackage{setspace}
\usepackage{bbm}
\usepackage{flushend}
\usepackage{dsfont}

\newcommand{\dP}{\mathrm{P}}
\newcommand{\dQ}{\mathrm{Q}}
\newcommand{\bPP}[1]{{\dP_{#1}}}
\newcommand{\bQQ}[1]{{\dQ_{#1}}}
\newcommand{\bPr}[1]{{\mathrm{P}}\left(#1\right)}

\newcommand{\unif}{\bPP {\mathtt{unif}}}

\newcommand{\cF}{{\mathcal F}}

\newcommand{\cX}{{\mathcal X}}
\newcommand{\cY}{{\mathcal Y}}
\newcommand{\cZ}{{\mathcal Z}}

\newcommand{\bF}{\mathbf{F}}

\newcommand{\ep}{\epsilon}

\usepackage{color}

\newtheorem{theorem}{Theorem}

\newtheorem{corollary}[theorem]{Corollary}
\newtheorem*{corollary*}{Corollary}

\newtheorem{lemma}[theorem]{Lemma}
\newtheorem*{lemma*}{Lemmas}

\theoremstyle{remark}
\newtheorem*{remark*}{Remark}
\newtheorem*{remarks*}{Remarks}

\theoremstyle{definition}

\newcommand{\ttlvrn}[2]{\left\| #1 - #2\right\|_1}
\newcommand{\KL}[2]{D\big( #1 \big\| #2\big)}
\newcommand{\CKL}[3]{D\big( #1 \big\| #2\, \big|\, #3\big)}
\newcommand{\MI}[2]{I\left(#1 \wedge #2\right)}
\newcommand{\CMI}[3]{I\left(#1 \wedge #2\mid #3\right)}

\newcommand{\cf}{{\it cf.}}
\newcommand{\ie}{{\it i.e.}}

\newcommand{\epdel}{(\ep, \delta)}
\newcommand{\Cepdel}{C_{\ep, \delta}}
\newcommand{\betepsn}{\beta_\ep(W,V,n)}
\newcommand{\betaepsdel}{\beta_{\ep+\delta+\eta}}
\newcommand{\betaepsdeln}{\beta_{\ep+\delta+\eta}(W,V,n)}

\usepackage[margin=54pt]{geometry}

\begin{document}
\IEEEoverridecommandlockouts

\title{Strong Converse for a Degraded Wiretap Channel via Active Hypothesis Testing}

\author{
\IEEEauthorblockN{Masahito Hayashi$^\ast$} 
\and
\IEEEauthorblockN{Himanshu Tyagi$^\dag$} 
\and
\IEEEauthorblockN{Shun Watanabe$^\ddag$} 
}

\maketitle

{\renewcommand{\thefootnote}{}\footnotetext{
\noindent$\ast$The Graduate School of Mathematics, Nagoya University, Japan,  and The Centre for
Quantum Technologies, National University of Singapore, Singapore. Email:masahito@math.nagoya-u.ac.jp

\noindent$^\dag$Information Theory and Applications (ITA) Center,
University of California, San Diego,
La Jolla, CA 92093, USA. Email: htyagi@eng.ucsd.edu

\noindent$^\ddag$Department of  Information Science and Intelligent Systems, 
University of Tokushima, Tokushima 770-8506, Japan, 
and Institute for Systems Research, University of Maryland, College Park,
MD 20742, USA. Email: shun-wata@is.tokushima-u.ac.jp
}}

\renewcommand{\thefootnote}{\arabic{footnote}}
\setcounter{footnote}{0}

\begin{abstract}
We establish an upper bound on the rate of 
codes for a wiretap channel with public feedback
for a fixed probability of error and secrecy
parameter. As a corollary, we obtain a
strong converse for the capacity of a degraded wiretap
channel with public feedback. Our converse proof is based on
a reduction of active hypothesis testing
for discriminating between two channels to 
coding for wiretap channel with feedback.
\end{abstract}

\section{Introduction}
We consider secure message transmission over a wiretap
channel $W:\cX\rightarrow \cY \times \cZ$ with noiseless, public feedback. For 
each transmission $x\in \cX$ over $W$,  the receiver 
observes a random output $Y \in \cY$ and an eavesdropper observes a correlated
side-information $Z\in \cZ$, with probability $W(Y, Z|x)$. Furthermore, the receiver 
can send a feedback to the transmitter over a noiseless channel. However, the feedback
channel is public and any communication sent over it is available to the eavesdropper.
The transmitter seeks to send a message $M$ to the receiver without revealing it to the
eavesdropper. For a given probability of error $\ep$ and a given secrecy parameter $\delta$, 
what is the maximum possible rate $\Cepdel$ of a transmitted message?

For a degraded wiretap channel $W$ with no feedback,
the wiretap capacity $C = \inf_{\ep, \delta}\Cepdel$ was established in the seminal
work of Wyner \cite{Wyn75ii} where it was shown that
\[
C = \max_{\bPP X}\CMI XYZ.
\] 
The capacity of a general wiretap channel was established in \cite{CsiKor78}. 
Extensions to wiretap channels with general statistics were considered in
\cite{Hay06}.
The model with feedback considered here was introduced in \cite{Cheong76}
where it was noted
that the availability of a noiseless feedback can enable positive rates of transmission over
a wiretap channel with zero capacity (see, also, \cite{Mau93}). However, the wiretap capacity with feedback remains unknown
in general; $\max_{\bPP X}\CMI XYZ$ constitutes an upper bound on it.

In this paper, we establish a {\it strong version} of this bound and show that 
for $\ep+\delta<1$
\[
\Cepdel \le \max_{\bPP X}\CMI XYZ,
\]
thereby characterizing $\Cepdel$ for all $0<\ep,\delta<1$ for a degraded wiretap channel.
A partial strong converse for a degraded wiretap channel
was established in \cite{MorganW14} for a restricted range of $\ep, \delta$. Another
strong converse for a degraded wiretap channel for
the case when $\delta\rightarrow 0$ was established, concurrently to this work, in 
\cite{TanB14}. In this work, we show a strong converse for all values of $\ep$ and $\delta$.

Our proof relies on a slight modification of a recent reduction of hypothesis testing to secret key agreement
shown in \cite{TyaWat14, TyaWat14ii}. Specifically, we show that a wiretap channel code yields an active hypothesis
test for distinguishing between two channels \cite{Hay09ii}. Consequently, the rate of a wiretap code is bounded above
by the rate of the optimum exponent of the probability of error of type II for discriminating a channel $W$ from 
another channel $V$
such that $V(y,z|x) = V_2(z|x)V_1(y|z)$, given that the probability of error of type I is less than $\ep+\delta$.
This gives an upper bound on the length of a wiretap code, which leads to the strong converse upon using the characterization
of the optimal exponent for channel discrimination derived in \cite{Hay09ii}.
This approach is along the lines of {\it meta-converse} of \cite{PolPooVer10}, where a reduction of hypothesis testing 
to channel coding was used to establish a finite-blocklength converse for the channel coding problem
(see, also, \cite{Nagaoka01} and \cite[Section 4.6]{Hayashi06}).

Our main result is given in the next section. Section \ref{s:hypothesis_testing} and \ref{s:secret_key} contains a review 
of relevant results in binary hypothesis testing and secret key agreement, respectively. The final section contains a proof
of our main result.

\section{Main result}\label{s:main_result}
We describe a generalization of the classic wiretap channel coding
problem \cite{Wyn75ii, CsiKor78} that was considered in \cite{Cheong76, Mau93, AhlCsi93},
where, in addition to transmitting over the wiretap channel, the terminals
can communicate using a noiseless, public feedback channel from the receiver to
the transmitter.

A wiretap code for a discrete\footnote{The restriction to discrete
alphabet is cosmetic. Our results apply to channels with continuous alphabet.
In particular, our strong converse holds for the Gaussian wiretap channel \cite{CheHel78}.} 
memoryless wiretap channel $W:\cX\rightarrow \cY\times \cZ$ with feedback
consists of (possibly randomized)
encoder mappings $e_t: \{1, ...,N\} \times \cF^{t} \rightarrow \cX$, $1\le t\le n$, feedback mappings
$f_t: \cY^t \rightarrow \cF$, $0\le t\le n-1$, and a decoder $d: \cY^n \rightarrow \{1, ..., N\}$. 
For a random message $M \sim \mathtt{unif}\{1,..., N\}$, the protocol begins with 
a feedback $F_0$ from the receiver at $t=0$. Subsequently, at each time
instance $1\le t\le n-1$ the transmitter sends $X_t = e_t(M, F^{t-1})$
and the channel outputs $(Y_t, Z_t)$ with probability $W(Y_t, Z_t| X_t)$. 
The receiver observes $Y_t$ and sends feedback $F_t = f_t(Y^t)$, and the eavesdropper
observes $Z_t$. The protocol stops with a final
transmission $X_n  = e_n(M, F^{n-1})$ over the channel and the subsequent decoding $\hat M = d(Y^n)$
by the receiver. We denote by $\bF$ the overall feedback communication $F_0, ..., F_{n-1}$.

The mappings $(\{e_t\}_{t=1}^n, \{f_t\}_{t=0}^{n-1}, d)$ constitute an $(N,n, \ep, \delta)$
wiretap code if 
\[
\bPr {M\neq \hat M}\le \ep,
\] 
and 
\[
\ttlvrn {\bPP {MZ^n \bF}}{\bPP M\times \bPP{Z^n\bF}}\le \delta,
\]
where $\ttlvrn \dP \dQ$ denotes the variation distance between $\dP$ 
and $\dQ$ given by
\[
\ttlvrn \dP \dQ  = \frac 12 \sum_x|\dP(x) - \dQ(x)|.
\] 
A rate $R>0$ is $\epdel$-achievable if
there exists an $(\lfloor 2^{nR}\rfloor, n ,\ep, \delta)$ wiretap code
for all $n$ sufficiently large. The
$\epdel$-{\it wiretap capacity} $\Cepdel$ is the supremum of all 
$\epdel$-achievable rates.

Our main result in an upper bound on $\Cepdel$

\begin{theorem}\label{t:main_result}
For $0\le \ep, \delta$ with $\ep+\delta< 1$, the $\epdel$-wiretap capacity is bounded above as
\[
\Cepdel \le \max_{\bPP X}\CMI XYZ.
\]
\end{theorem}  
For the special case of a degraded wiretap channel $W$ with 
$W(y,z|x) = W_1(y|x)W_2(z|y)$, Theorem \ref{t:main_result}
yields a {\it strong converse} for wiretap capacity.
\begin{corollary}
For a degraded wiretap channel $W$, 
\[
\Cepdel = \begin{cases}
\displaystyle \max_{\bPP X} \CMI XYZ, &\quad 0< \ep < 1-\delta,
\\
\displaystyle \max_{\bPP X} \MI XY, &\quad 1-\delta \le \ep <1.
\end{cases}
\]
\end{corollary}
{\it Proof.} 
For $0< \ep <1-\delta$, the result is an immediate corollary of Theorem \ref{t:main_result} 
and \cite{Wyn75ii}\footnote{While the secrecy criterion in \cite{Wyn75ii} is different
from variational secrecy required here, the achievability result for the latter 
follows from the results in \cite{Csi96, Hay06}.}. 
For $1-\delta\le \ep <1$, the converse follows from the strong converse
for the capacity of a DMC with feedback (\cf{} \cite{PolVer10}). Moving to the proof of achievability, it suffices to 
restrict to $\ep+\delta=1$. For this case, achievability follows by randomizing between
an $(\ep_n, 1)$ wiretap code, $\ep_n \rightarrow 0$ as $n\rightarrow \infty$, and a $(1, 0)$ wiretap code --
the randomizing bit is communicated as the public feedback $F_0$ by the receiver\footnote{Alternatively, the sender can 
transmit the randomizing bit over the wiretap channel with neglible rate loss.}
\qed

As a preparation for the proof of Theorem \ref{t:main_result} given in Section~\ref{s:proof_main_result},
we review some results in hypothesis testing and secret key agreement in the next two sections.
\section{Hypothesis testing}\label{s:hypothesis_testing}
Consider a simple binary hypothesis testing problem with null hypothesis $\mathrm{P}$ and alternative hypothesis 
$\mathrm{Q}$, where $\mathrm{P}$ and $\mathrm{Q}$ are distributions on the same alphabet ${\cal X}$. Upon observing 
a value $x\in \cX$, the observer needs to decide if the value was generated by the distribution $\bPP{}$ or the 
distribution $\mathrm{Q}$. To this end, the observer applies a stochastic test $\mathrm{T}$, which is a conditional 
distribution on $\{0,1\}$ given an observation $x\in \cX$. When $x\in \cX$ is observed, the test $\mathrm{T}$ 
chooses the null hypothesis with probability $\mathrm{T}(0|x)$ and the alternative hypothesis with probability 
$T(1|x) = 1 - T(0|x)$. For $0\leq \ep<1$, denote by $\beta_\ep(\mathrm{P},\mathrm{Q})$ the infimum of the probability 
of error of type II given that the probability of error of type I is less than $\ep$, i.e.,
\begin{eqnarray}
\beta_\ep(\mathrm{P},\mathrm{Q}) := \inf_{\mathrm{T}\, :\, \mathrm{P}[\mathrm{T}] \ge 1 - \ep} \mathrm{Q}[\mathrm{T}],
\nonumber
\end{eqnarray}
where 
\begin{eqnarray*}
\mathrm{P}[\mathrm{T}] &=& \sum_x \mathrm{P}(x) \mathrm{T}(0|x), \\
\mathrm{Q}[\mathrm{T}] &=& \sum_x \mathrm{Q}(x) \mathrm{T}(0|x).
\end{eqnarray*}
The following result credited to Stein characterizes the optimum exponent of
$\beta_\ep(\mathrm{P}^n,\mathrm{Q}^n)$ where $\dP^n  = \dP \times ...\times \dP$
and $\dQ^n = \dQ\times ... \times \dQ$.
\begin{lemma}\emph{(\cf{} \cite[Theorem 3.3]{Kul68})} For every $0< \ep <1$, we have 
\begin{eqnarray}
\lim_{n\to\infty} - \frac{1}{n} \log \beta_\ep(\mathrm{P}^n,\mathrm{Q}^n) = D(\mathrm{P} \| \mathrm{Q}),
\nonumber
\end{eqnarray}
where $D(\mathrm{P}\|\mathrm{Q})$ is the Kullback-Leibler divergence  given by
\begin{eqnarray*}
D(\mathrm{P}\|\mathrm{Q}) = \sum_{x\in \cX} \mathrm{P}(x) \log \frac{\mathrm{P}(x)}{\mathrm{Q}(x)},
\end{eqnarray*}
with the convention $0\log(0/0) = 0$.
\end{lemma}

Next, we review a problem of active hypothesis testing where the distribution
at each instance is determined by a prior action. Specifically, 
given two 
DMCs $W:\cX\rightarrow \cY$ and $ V:\cX\rightarrow \cY$, we seek to design
a transmission-feedback scheme such that by observing the channel inputs, channel outputs,
and feedback we can determine if the underlying channel is $W$ or $V$. Formally, an $n$-length active hypothesis
test consist of (possibly randomized) encoder mappings $e_t:\cF^t\rightarrow \cX$, $1\le t\le n$,
feedback mappings  $f_t:\cY^t \rightarrow \cF$, $0\le t\le n-1$, and a conditional distribution $T$ on $\{0,1\}$ 
given $X^n, Y^n, \bF$. On observing $X^n, Y^n, \bF$, we detect the null hypothesis $W$ with probability $T(0| X^n, Y^n, \bF)$
and alternative hypothesis $V$ with probability $T(1| X^n, Y^n, \bF)$. Analogous to $\beta_\ep(\dP, \dQ)$,
the quantity $\beta_\ep(W,V,n)$, for $0\le \ep<1$, is 
the infimum of the probability of error of type II over all $n$ length active hypothesis tests for null hypothesis $W$
and alternative hypothesis $V$ such that the probability of error of type I
is no more than $\ep$.

The following analogue of Stein's lemma for active hypothesis testing was established in \cite{Hay09ii} (see, also, \cite{PolVer10}).
\begin{theorem}[\cite{Hay09ii}]\label{t:active_HT}
For $0< \ep <1$,
\begin{align*}
\lim_n - \frac 1n \log \betepsn &= \max_{\bPP X} \CKL WV{\bPP X}
\\
&= \max_x \KL {W_x}{ V_x},
\end{align*}
where $W_x$ and $V_x$, respectively, denote the $x$th row of $W$ and $V$.
\end{theorem}
Remarkably, the exponent above is achieved without any feedback, \ie, while feedback is available,
it does not help to improve the asymptotic exponent of $\betepsn$.
\section{Secret key agreement}\label{s:secret_key}
In this section, we review two party secret key (SK) agreement where
parties observing random variables $X$ and $Y$
communicate interactively over a public channel to agree on a SK 
that is concealed from an eavesdropper with access to the communication
and a side-information $Z$. 

Formally, the parties communicate using an interactive communication $\bF = F_1, ..., F_r$
where $F_1 = F_1(X), F_2 = F_2(Y, F_1), F_3 = F_3(X, F^2)$, $F_4 = F_4(Y, F^3)$ and so on. 
A random variable 
$K = K(X,\bF)$ constitutes an $\epdel$-SK if there exists $\hat K  = \hat K(Y, \bF)$ such that
\begin{align}
\bPr {K\neq \hat K}\le \ep,
\nonumber
\end{align}
and 
\begin{align}
\ttlvrn {\bPP {KZ\bF}}{\unif \times \bPP {Z\bF}}\le \delta.
\nonumber
\end{align}
The following upper bound on the number of values $k$
taken by an $\epdel$-SK $K$ was shown in \cite{TyaWat14, TyaWat14ii}:
\[
\log k \le - \log \betaepsdel(\bPP{XYZ}, \bQQ{XYZ}) + 2\log \frac 1\eta,
\]
for all $0< \eta < 1-\ep -\delta$, and all $\bQQ{XYZ} = \bQQ{X|Z} \bQQ{Y|Z} \bQQ{Z}$.
Underlying the proof of this bound is an intermediate reduction argument 
in \cite[Lemma 1]{TyaWat14}
that relates SK agreement to hypothesis testing. We recall this result below.
\begin{theorem}[\cite{TyaWat14, TyaWat14ii} ]\label{t:SK_reduction}
For $0\le \ep, \delta, \ep+\delta<1$, let random variables $K,\hat K$, and $Z$ be such 
that $\bPr {K \neq \hat K}\le \ep$ and 
\[
\ttlvrn {\bPP{KZ}}{\unif\times\bPP{Z }}\le \delta,
\]
where $\unif$ denotes a uniform distribution on $k$ values. Then,
for every $0<\eta< 1-\ep -\delta$ and every $\bQQ {K\hat K Z} = \bQQ{K\mid Z}\bQQ{\hat K \mid Z}\bQQ Z$,
\[
\log k \le - \log \betaepsdel(\bPP{K\hat K Z},\bQQ{K \hat KZ}) + 2\log \frac 1\eta.
\]
\end{theorem}
\section{Proof of main result}\label{s:proof_main_result}
We present a converse result that applies for every fixed
$n$ and is asymptotically tight, giving the strong converse
result of Theorem \ref{t:main_result}.

\begin{theorem}\label{t:wiretap_converse}
For $0\le \ep, \delta, \ep+\delta< 1$, given an $(N, n, \ep, \delta)$-wiretap
code, we have 
\[
\log N \le -\log \betaepsdeln + 2\log \frac 1\eta,
\]
for all $0< \eta < 1-\ep-\delta$ and all channels $V: \cX \rightarrow \cY \times \cZ$
such that $V(y, z|x) = V_2(z|x)V_1(y|z)$.
\end{theorem}
{\it Proof of Theorem \ref{t:main_result}.} Theorem \ref{t:main_result} follows form Theorems \ref{t:wiretap_converse}
and \ref{t:active_HT} upon noting that for $W(y, z|x) = W_2(z|x)W_1(y|z, x)$ 
\begin{align*}
&\hspace{-1em}\min_V \max_{\bPP X}\,\CKL W V {\bPP X} 
\\
&= \min_{V_1}\max_{\bPP X}\, \CKL {W_1}{V_1}{\bPP X W_2}
\\
&= \max_{\bPP X} \min_{V_1}\,\CKL {W_1}{V_1}{\bPP X W_2}
\\
&= \max_{\bPP X}\, \CKL {\bPP {Y\mid ZX}}{\bPP{Y\mid Z}}{\bPP{ZX}}
\\
&= \max_{\bPP X}\,\CMI XYZ,
\end{align*}
where $\bPP{XYZ}$ is given by $\bPP XW$.
\qed

We need the following result to prove Theorem~\ref{t:wiretap_converse}.
\begin{lemma}\label{l:conditional_independence}
For a wiretap channel $V:\cX\rightarrow \cY \times \cZ$ such that $V(y,z|x) = V_2(z|x)V_1(y|z)$,
a random message $M$, and a wiretap code, let $\hat M =d(Y^n)$ and $\bF$ be the corresponding feedback.
Then, the induced distribution 
$\bQQ{M \hat M Z^n \bF}$ satisfies factorization condition
\[
\bQQ{M \hat M\mid Z^n \bF} = \bQQ{M\mid Z^n\bF}\times\bQQ{\hat M\mid Z^n \bF}.
\]
\end{lemma}

{\it Proof of Lemma~\ref{l:conditional_independence}.} Denote by $U_x$ and $U_y$, respectively, the local randomness 
at the transmitter and the receiver, and by $F^t$ the feedback $(F_0, ..., F^t)$. 
Thus, the encoder mapping $e_t$ is a (deterministic) function of $(M, U_x, F^{t-1})$
and the feedback mapping $f_t$ is a (deterministic) function of $(Y^t, U_y)$. 
The proof entails a repeated application of the fact that conditionally independent random variables
remain so when conditioned additionally on an interactive communication (cf.~\cite{TyaNar13ii})
and is completed by induction. Specifically, note first that
$\bQQ{MU_xU_y \mid F_0} = \bQQ{MU_x\mid F_0}\bQQ{U_y \mid F_0}$
since $(M,U_x)$ and $U_y$ are independent and $F_0$ is an interactive communication.
Under the induction hypothesis 
\begin{align*}
&\hspace{-1em}\bQQ{MU_xX^{t-1}U_yY^{t-1}\mid Z^{t-1}F^{t-1}} 
\\
&=\bQQ{MU_xX^{t-1}\mid Z^{t-1} F^{t-1}} \bQQ{U_yY^{t-1}\mid Z^{t-1} F^{t-1}},
\end{align*}
we get
\begin{align*}
&\hspace{-1em}\CMI{M,U_x,X^t}{U_y,Y^t}{Z^t, F^{t-1}} 
\\
&= \CMI{M,U_x,X^t}{U_y,Y^{t-1}}{Z^t, F^{t-1}}
\\
&\leq \CMI{M,U_x,X^t}{U_y,Y^{t-1}}{Z^{t-1}, F^{t-1}}
\\
&= \CMI{M,U_x,X^{t-1}}{U_y,Y^{t-1}}{Z^{t-1}, F^{t-1}}
\\
&=0,
\end{align*}
where the first equality and inequality follow since $Y_t$ and $Z_t$, respectively, are outputs of $V_1$
for input $Z_t$ and $V_2$ for input $X_t$, and the second equality holds since $X_t = e_t(M, U_x, F^{t-1})$,
which completes the proof.
\qed

{\it Proof of Theorem \ref{t:wiretap_converse}.} Given an $(N, n ,\ep, \delta)$ wiretap code, a message $M\sim \mathtt{unif}\{1,...,N\}$
and its decoded value $\hat M = d(Y^n)$ satisfy the conditions for Theorem~\ref{t:SK_reduction}
with $K = M, \hat K = \hat M,$ and $Z = (Z^n, \bF)$. Letting $\bQQ {M\hat M Z^n\bF}$ be the distribution on 
$(M,\hat M, Z^n, \bF)$ when the underlying channel is $V$, by Lemma~\ref{l:conditional_independence}
and Theorem~\ref{t:SK_reduction} we get
\[
\log N \le - \log \betaepsdel(\bPP {M\hat M Z^n\bF}, \bQQ {M\hat M Z^n\bF}) + 2\log \frac 1\eta.
\]
Note that a test for the simple binary hypothesis testing problem for $\bPP {M\hat M Z^n\bF}$
and $\bQQ {M\hat M Z^n\bF}$ along with the wiretap code constitutes an active hypothesis test for
$W$ and $V$. Therefore,
\begin{align*}
&\hspace*{-1em}- \log \betaepsdel(\bPP {M\hat M Z^n\bF}, \bQQ {M\hat M Z^n\bF}) 
\\
&\le - \log \betaepsdeln,
\end{align*}
which completes the proof.\qed

\section*{Acknowledgements}
MH is partially supported by a MEXT Grant-in-Aid for Scientific Research (A) No. 23246071. 
MH is also partially supported by the National Institute of
Information and Communication Technology (NICT), Japan.
The Centre for Quantum Technologies is funded by the Singapore Ministry
of Education and the National Research Foundation as part of
the Research Centres of Excellence programme.
\bibliography{IEEEabrv,references}
\bibliographystyle{IEEEtranS}

\end{document}